\documentclass[prl,twocolumn,showpacs,amsmath,amssymb,floatfix,superscriptaddress]{revtex4-1}
\usepackage{bm}% bold math
\usepackage{color}
\usepackage{graphicx}
\usepackage[normalem]{ulem}
\usepackage[makeroom]{cancel}
\usepackage{multirow}

\usepackage{hyperref}
\hypersetup{
  breaklinks = {true},
  citecolor = {blue},
  colorlinks = {true},
  linkcolor = {red},
  urlcolor = {blue},
}

\graphicspath{{../Figs/}}

\begin{document}

\title{Orbital-spin Locking and its Optical Signatures in Altermagnets}

\author{Marc Vila}
\email{mvilatusell@lbl.gov}
\affiliation{Department of Physics, University of California, Berkeley, California 94720, USA}
\affiliation{Materials Sciences Division, Lawrence Berkeley National Laboratory, Berkeley, California 94720, USA}
%\affiliation{Molecular Foundry, Lawrence Berkeley National Laboratory, Berkeley, California 94720, USA}

\author{Veronika Sunko}
\affiliation{Department of Physics, University of California, Berkeley, California 94720, USA}
\affiliation{Materials Sciences Division, Lawrence Berkeley National Laboratory, Berkeley, California 94720, USA}

\author{Joel E. Moore}
\affiliation{Department of Physics, University of California, Berkeley, California 94720, USA}
\affiliation{Materials Sciences Division, Lawrence Berkeley National Laboratory, Berkeley, California 94720, USA}

\date{\today}

\begin{abstract}

Altermagnets, magnetic materials with zero magnetization and spin-split band structure, have gained tremendous attention recently for their rich physics and potential applications. Here, we report on a microscopic tight-binding model that unveils a unique coupling between orbitals and spins in $d$-wave altermagnets which gives rise to momentum-dependent and spin-selective optical absorption. This coupling promotes the controlled optical excitation of up or down spins depending on the polarization direction of linearly polarized light. Such an effect originates from the coupling of orbitals to the sublattice degree of freedom through the crystal field, which is then coupled to spins through the antiferromagnetic interaction. Our crystal field analysis, which is general to any type of altermagnet, helps understand the onset of altermagnetism from a microscopic point of view, and we use our results to propose clear magneto-optical signatures of our predictions. Our findings shine light on the interplay between orbitals and spins in altermagnets, thus paving the way towards novel orbitronic and opto-spintronic devices.

\end{abstract}

\maketitle

\textit{Introduction ---} Recent developments in the study of spin splitting in antiferromagnetic materials \cite{Noda2016, Smejkal2020, Naka2019, Ahn2019, Hayami2019, Yuan2020, Hayami2020, Samanta2020, Gonzalez2021, Feng2022, Yuan2021, Hayami2020B, Shao2021, Smejkal2022C} has unveiled a type of magnetic materials with distinct symmetry classification dubbed altermagnets (ALTMs) \cite{Smejkal2022,Smejkal2022B}. These materials possess total zero magnetization, akin to antiferromagnets, but display properties that are usually associated with ferromagnets, such as spin-split band structures or the anomalous Hall effect \cite{Smejkal2020, Smejkal2022B, Smejkal2022Rev}. In ALTMs, the symmetry operation that combines with time-reversal to connect sublattices with opposite magnetization is a rotation instead of a translation or inversion, and the type of rotation serves to classify the type of altermagnetism; for example, a four-fold (six-fold) rotation denotes a $d$-wave ($g$-wave) ALTM \cite{Smejkal2022,Smejkal2022B, McClarty2024}. This symmetry appears in many of the properties of ALTMs, such as in relating the sign of the spin splitting across the Brillouin zone, and can therefore be used as a fingerprint of this magnetic phase.

Optical methods are a well-established and versatile tool to both detect and control antiferromagnetic order \cite{Kirilyuk2010,Saidl2017, Nemec2018}. However, although many properties of ALTMs have already been actively investigated \cite{Smejkal2023, Liu2024, Gonzalez2021, Bai2023, Wang2024, Liao2024, Papaj2023, Ouassou2023, Ghorashi2024, Mazin2023, Fernandes2024, Aoyama2024, Chakraborty2024, Bhowal2024, Zhou2024, McClarty2024, Fernandes2024, Antonenko2024} and many materials have been experimentally identified \cite{Krempasky2024, Zhu2024, Lee2024, Osumi2024, Fedchenko2024, Lin2024, Reimers2024, Yang2024, Zeng2024CrSb, Reichlova2024, Regmi2024}, the study of their optical properties has only started to receive attention recently \cite{Rao2024, Kimel2024, HarikiRuO2, Hariki2024, Adamantopoulos2024, HarikiMnF2, Weber2024}. Yet, a microscopic model capable of unveiling optical responses of ALTMs in a transparent manner is currently lacking. The focus of many minimal models is on capturing the spin splitting only, neglecting in this way the multi-orbital nature of real materials \cite{Smejkal2022C, Brekke2023, Fang2024, Antonenko2024, Attias2024, Rao2024}, which is a necessary ingredient to study optical phenomena. Furthermore, some theoretical works have predicted optical fingerprints of ALTMs, but they are case-specific studies of individual materials \cite{HarikiRuO2, Hariki2024, Adamantopoulos2024, HarikiMnF2}. Overall, a general model including multi-orbital effects capable of describing and predicting optical phenomena in altermagnets is needed.

In this Letter, by developing a multi-orbital tight-binding model, we demonstrate a unique coupling between orbitals and spins in \textit{d}-wave altermagnets that gives rise to momentum-dependent and spin-selective optical absorption. As a result, linearly polarized light selectively excites up or down spins. Motivated by this effect, we propose a magneto-optical experiment to detect not only the presence of altermagnetism, but also the N\'eel vector orientation. The explicit inclusion of crystal field (CF) effects in the model helps understand this phenomena from the coupling between orbital, sublattice and spins degrees of freedom. Our results not only demonstrate unique optical manipulation of orbitals and spins in ALTMs but also clarify the fundamental role of crystal field in these novel magnetic materials. 

\textit{Microscopic model ---} Our goal is to formulate a tight-binding model capturing orbital physics and altermagnetism. It is understood \cite{Smejkal2022,Smejkal2022B,Yuan2021B, Yuan2023} that nonmagnetic ligands surrounding the magnetic ions are crucial to bring about the necessary symmetries for altermagnetism {in many materials. Importantly, these ligands arrange differently around magnetic ions with opposite magnetization, and microscopically, they produce a crystal field with the same strength but unequal orientation in the two sublattices (both the ligands' position and the CF orientation in the two sublattices are related by the aforementioned rotation symmetry). It is then natural to consider crystal field effects in our model, and to gain that insight, one must look at the local symmetry of each sublattice. 

\begin{figure}[bt]
    \centering
    \includegraphics[width=0.48\textwidth]{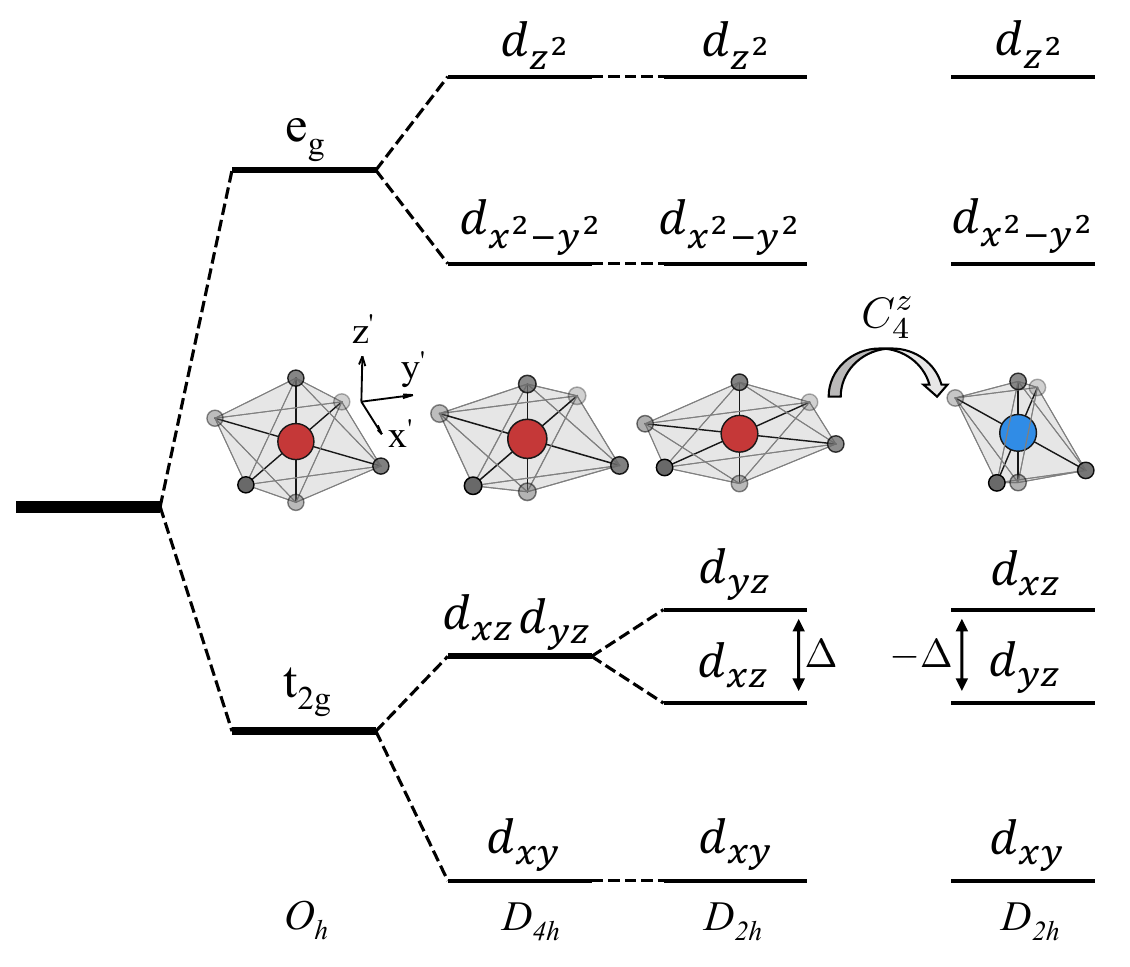}
\caption{Representative schematics of the local crystal field splitting of a \textit{d}-wave altermagnet. Grey atoms are the nonmagnetic ligands and red and blue atoms are the magnetic ions with opposite magnetization that belong to opposite sublattices. The $e_g$ and $t_{2g}$ orbitals of an octahedral coordination split upon compression, resulting in a $D_{4h}$ point group. Further distortions lower the symmetry to a $D_{2h}$ group. Lastly, a four-fold rotation ($C_4^z$) transforms the crystal field from one sublattice to the other, swapping the crystal field sign that splits the $d_{xz}$ and $d_{yz}$ orbitals.}
    \label{fig_fig1}
\end{figure}

Fig. \ref{fig_fig1} shows a representative example of how the various distortions of the local coordination environment of the transition metal atom affect the atomic energy levels in a \textit{d}-wave ALTM such as MnF$_2$ or RuO$_2$ \cite{Occhialini2021}. A transition metal atom is surrounded by ligands, producing an octahedral $O_h$ crystal field splitting between $e_g$ and $t_{2g}$ orbitals. Distorting this octahedron by compressing the bonds along the local $z^\prime$ direction breaks all orbital degeneracies except that between $d_{xz}$ and $d_{yz}$ orbitals. Finally, distorting the bond angles in the $x^\prime y^\prime$ plane reduces the high-symmetry axis to a two-fold rotation only which splits the $d_{xz}$ and $d_{yz}$ orbitals. Each symmetry breaking can be associated with a different crystal field term \cite{Occhialini2021, suppmat}, and importantly, the sign of the splitting between $d_{xz}$ and $d_{yz}$ orbitals is changed when applying the altermagnetic rotational symmetry. In other words, while the local symmetry of the magnetic ions tells us about the form of the CF, the global symmetry rotation produces a sublattice-dependent CF. Although we have illustrated this on the example of a $d$-wave altermagnet (such as RuO$_2$, MnF$_2$, V$_2$Se$_2$O or RuF$_4$ \cite{Smejkal2022, Higuchi2016, Ma2021, Milivojevic2024}), the same procedure can be applied to any kind of lattice by inspecting the local crystal field and observing how the energy levels of the orbitals transform under the rotation symmetries connecting the sublattices.

Next, we formulate a tight-binding model in an antiferromagnetic lattice that captures the CF physics mentioned above. We assume that the competition of kinetic energy, electronic correlations and CF have stabilized an antiferromagnetic order that can be effectively described by a single-particle model. For concreteness, we assume that the magnetic anisotropy energy promotes a preferred direction of the N\'eel vector, e.g., along the $z$ direction. We use $d_{xz}$ and $d_{yz}$ orbitals for our model, motivated by the crystal field analysis from Fig. \ref{fig_fig1}, and supported by \textit{ab initio} calculations on RuO$_2$ \cite{Guo2023}. Thus, our model is composed of two sublattices (A and B, see Fig. \ref{fig_fig2}(a)), orbital and spin degrees of freedom. We use a two-dimensional square lattice for simplicity but the extension to three dimensions is straightforward. The hopping terms are given by the Slater-Koster parametrization \cite{Slater1954}, and the tight-binding Hamiltonian obtained reads
\begin{align}
\mathcal{H} &= \mathcal{H}_{0} + \mathcal{H}_{CF} + \mathcal{H}_{AFM} \\
\mathcal{H}_{0} &= \sigma_x \left( T_x \cos{k_x a}  + T_y \cos{k_y a} \right) s_0 \label{eq_hopping} \\
\mathcal{H}_{CF} &= \Delta \, \sigma_z \tau_z s_0 \label{eq_CF} \\
\mathcal{H}_{ex} &= m \, \sigma_z \tau_0 s_z. \label{eq_M}
\end{align}
Here, $\mathcal{H}_{0}$ describes the nearest-neighbor hopping between sites, $\mathcal{H}_{CF}$ is the CF that splits the $d_{xz}$ and $d_{yz}$ orbitals and has opposite signs in different sublattices \cite{suppmat} and $\mathcal{H}_{AFM}$ is an antiferromagnetic term that induces an effective exchange interaction that also changes sign with sublattice \cite{Smejkal2020}. $T_x$ and $T_y = {C_4^{z}}^\dagger T_x C_4^z$ are the nearest-neighbor hopping matrices along the $x$ and $y$ directions, respectively, and are expressed in the basis of $d_{xz}$ and $d_{yz}$ orbitals as $T_x = \text{diag}(V_\pi, V_\delta)$ and $T_y = \text{diag}(V_\delta, V_\pi)$, with $V_\pi$ and $V_\delta$ being the typical Slater-Koster bond integrals and $C_4^z$ a four-fold rotation about the $z$ axis. $\Delta$ is the CF strength, $m$ is the exchange interaction, and $\sigma$, $\tau$ and $s$ are Pauli matrices acting on the sublattice, orbital and spin spaces, respectively. This model preserves the symmetries of the $D_{4h}$ point group, including inversion symmetry. The model can straightforwardly be extended to include local spin-orbit coupling $\mathcal{H}_{SOC} \propto \vec{L} \cdot \vec{s}$ ($\vec{L}$ is the angular momentum operator). We chose not to do that here, since spin-orbit coupling is expected to be weaker than the onsite crystal field energy \cite{Occhialini2021}, and therefore should not significantly change the local picture shown in Fig. \ref{fig_fig1}. We note that SOC is required to observe other signatures of altermagnetism, such as the anomalous Hall effect and the Kerr effect \cite{Feng2022, Gonzalez2023, Hariki2024}, but is not required for the magneto-optical effects we predict below.

In Figs. \ref{fig_fig2}(c-k) we plot the band structure for a range of parameters along a path between the high-symmetry points shown in Fig. \ref{fig_fig2}(b), and we color the bands by their sublattice, orbital and spin projection. We start by setting $\Delta = 0$, $m = 0$ and $V_\pi > V_\delta = 0$ (Figs. \ref{fig_fig2}(c-e)). Between the $M_1 - \Gamma$ and $M_2 - \Gamma$ paths, which are parallel to the $k_x$ and $k_y$ axes, respectively, there are four bands (each doubly spin degenerate), while for the rest of the $k$-paths there are two bands, each four-fold degenerate. The sublattice projection is zero and the spins are degenerate, but the doubly degenerate bands show differences in the orbital content. This originates because $V_\pi \neq V_\delta$: along the $k_x$ direction, $d_{xz}$ and $d_{yz}$ orbitals interact with strength $V_\pi$ and $V_\delta$, respectively and hence it is expected that their energies and dispersion to be different. Along $k_y$, the bonding type reverses and so does the orbital polarization. 
%\note{local symm dictates which orbitals have the sublattice dependent CF. global symmetry + coordinate axes tells you the wave and which k path the splitting is. for altm to occur, the orbitals at play in the local symmetry have the have different energies/hoppings along. in other words: local dictates the basis/orbitals, global dictates the lattices. basis + lattice is the only thing you need for slater koster.}

\begin{figure}[bt]
    \centering
    \includegraphics[width=0.48\textwidth]{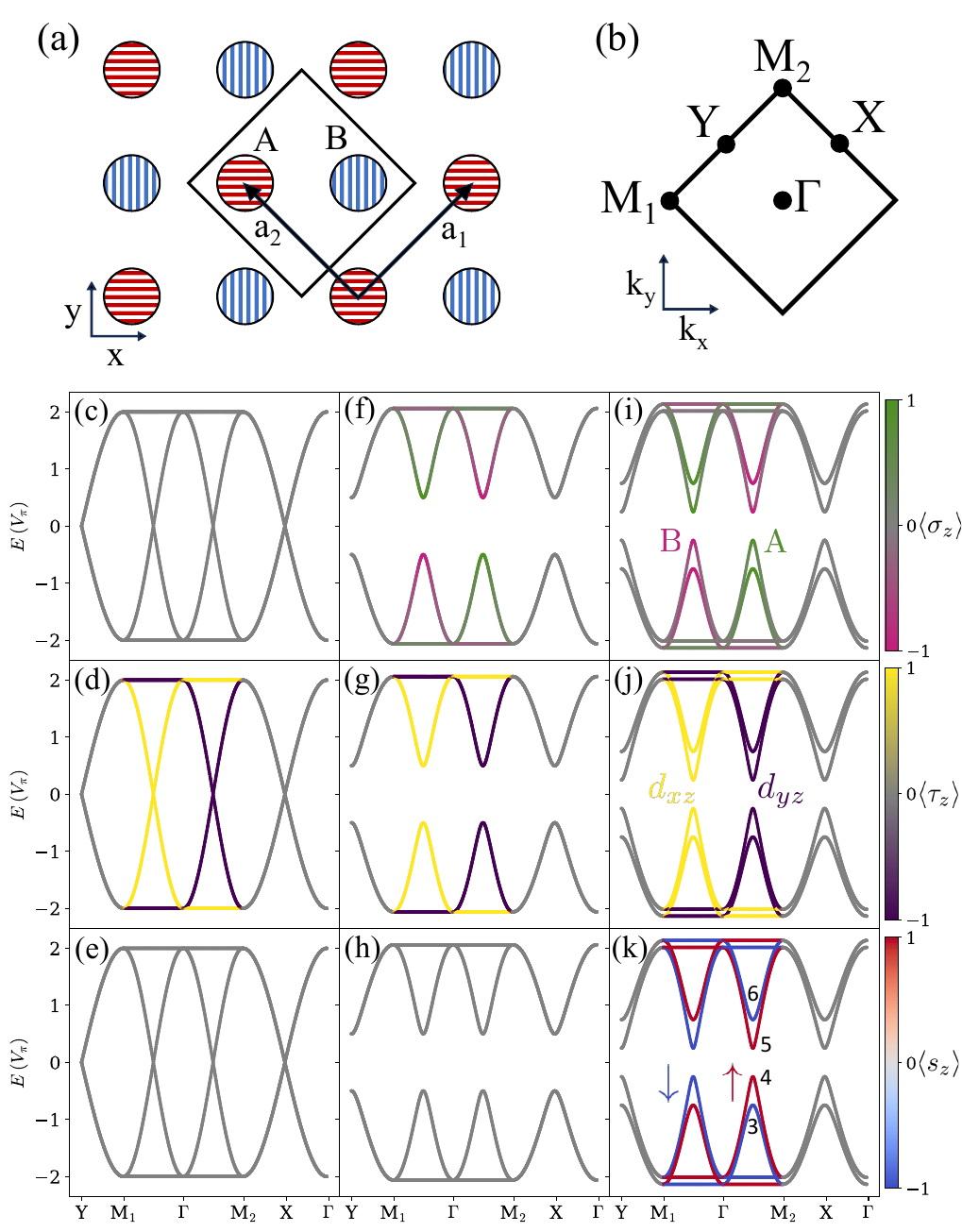}
\caption{(a) Real space lattice of our tight-binding model. Red and blue color denote up and down magnetization, and horizontal and vertical stripes indicate opposite directions of the local crystal field (i.e. horizontal (vertical) lines increase the energy of $d_{xz}$ ($d_{yz}$) orbitals). (b) Brillouin zone of the lattice in (a). (c-k) Band structure of the altermagnetic model. (c-e) Bands with only hopping terms $V_\pi = 1$, $V_\delta = 0$; (d-h) bands with added crystal field $\Delta = 0.5$; (i-k) bands with added exchange interaction $m = 0.25$ (in units of $V_\pi$). Top row (c, f, i) show the sublattice polarization, middle row (d, g, j) show the orbital polarization, bottom row (e, h, k) show the spin polarization.}
%$V_\pi = 1$, $\Delta = 1$, $m = 1.5$.
    \label{fig_fig2}
\end{figure}

We then add a nonzero $\Delta$ and plot the resulting bands and projections in Figs. \ref{fig_fig2}(f-h). This produces two main effects: a gap opening between the $M_1 - \Gamma$ and $M_2 - \Gamma$ paths and a nonzero sublattice polarization (the orbital and spin polarization remain the same). The former effect will be relevant later to study optical excitations and we now focus on the latter. According to Eq. \eqref{eq_CF}, the crystal field raises the energy of the $d_{xz}$ ($d_{yz}$) orbital in the A (B) sublattice while lowering its energy in the other sublattice. One can indeed see that pattern in the sublattice and orbital projections: bands with $d_{xz}$ character above the band gap are polarized on sublattice A while those below the gap reside in sublattice B (and the opposite holds for $d_{yz}$ orbital). 
%It is worth mentioning that the bands along the $M_1 - \Gamma$ and $M_2 - \Gamma$ paths are always degenerate: the local crystal field does act oppositely on each sublattice, but the global $C_4^zT$ connects them. 
The last ingredient needed to trigger altermagnetism is the antiferromagnetic term. Because the bands along the $M_1 - \Gamma$ and $M_2 - \Gamma$ paths have opposite sublattice polarization, their spin splitting is also opposite (see Eq. \eqref{eq_M}), which then gives rise to $d$-wave altermagnetism (Fig. \ref{fig_fig2}(k)). 

A few notes are in order at this point. One could view altermagnetism as a combination of antiferromagnetism and crystal field effects coupled through the sublattice degree of freedom. Nevertheless, this picture is incomplete, as the anisotropy of the electronic hoppings and bonding is also a necessary component. Indeed, we show in the supplemental material \cite{suppmat} that altermagnetism does not occur when $V_\pi = V_\delta$ even when both $\Delta$ and $m$ are nonzero. Although past models have considered anisotropy in the hopping terms \cite{Smejkal2022C, Brekke2023, Fang2024, Roig2024, Attias2024}, they described an effective spin-dependent hopping without referring to an underlying physical interpretation. Here, we explicitly describe such an anisotropic environment with $V_\pi$, $V_\delta$ and $\Delta$, which are clear and understood quantities in both physics and   chemistry. We also note that Ref. \cite{Antonenko2024} explained the anisotropic environment of the Lieb lattice as originating by projecting out the high-energy states coming from the nonmagnetic atoms. Finally, from the results of our model, \textit{several} spin splittings between different pairs of bands can be identified with unequal dependencies on the model parameters and different magnitudes at different $k$-points \cite{suppmat}. This enriches the spin-split picture of ALTMs obtained with more simplified models. 

\textit{Optical properties ---} In addition to the altermagnetic splitting, our model also reveals orbital polarization in the spin-split bands (Fig. \ref{fig_fig2}(j)). Such coupling between spin and orbitals opens the door to a variety of probes of spin by means of orbital physics. We note that the alternating spin splitting of altermagnets resembles the spin-valley locking of transition metal dichalcogenides \cite{Xiao2012, Smejkal2022B} (with the difference that here the splitting is even in momentum space), which are known to display rich optical properties. This motivates us to study optical transitions in our model, and, as shown below, they can provide unique signatures of altermagnetic order. 

We start by computing the optical matrix elements $|M_{\alpha}^{ji}(\bm{k})|^2$ between pair of bands $i,j$ with linearly polarized light ($\alpha=x,y$), with $M_{\alpha}^{ji}(\bm{k}) = \langle u_j(\bm{k})| \frac{1}{\hbar} \frac{\partial H}{\partial k_\alpha} \big|  u_i(\bm{k}) \rangle$. We plot in Fig. \ref{fig_fig3} the results for $i=4, j=5$, where the band indices are shown in Fig. \ref{fig_fig2}(k). The $k$-points corresponding to the optically-allowed transitions are quite localized in momentum space and are different between $x$ and $y$ polarization. For $x$ ($y$), the bands participating in the transition run mainly along the $k_x$ ($k_y$) direction since light polarized along $x$ ($y$) couples only to $d_{xz}$ ($d_{yz}$) orbitals (Fig. \ref{fig_fig2}(j)). These results have pronounced implications for the spins, because the orbitals are coupled to a specific spin. Indeed, the transition for bands $i=4, j=5$ is allowed because the spin is conserved (see right plot in Fig. \ref{fig_fig3}). In this manner, $x$-polarized light is able to selectively excite down spins, whereas $y$-polarized light only excites up spins. Moreover, changing the pair of bands to $i=3, j=6$ (i.e. changing the photon energy) reverses the orbital-spin locking and therefore the opposite spin is excited without changing the light polarization (not shown). We note that the sublattice projection changes between valence and conduction states, indicating that the transition is a charge transfer excitation.

\begin{figure}[t!]
    \centering
    \includegraphics[width=0.48\textwidth]{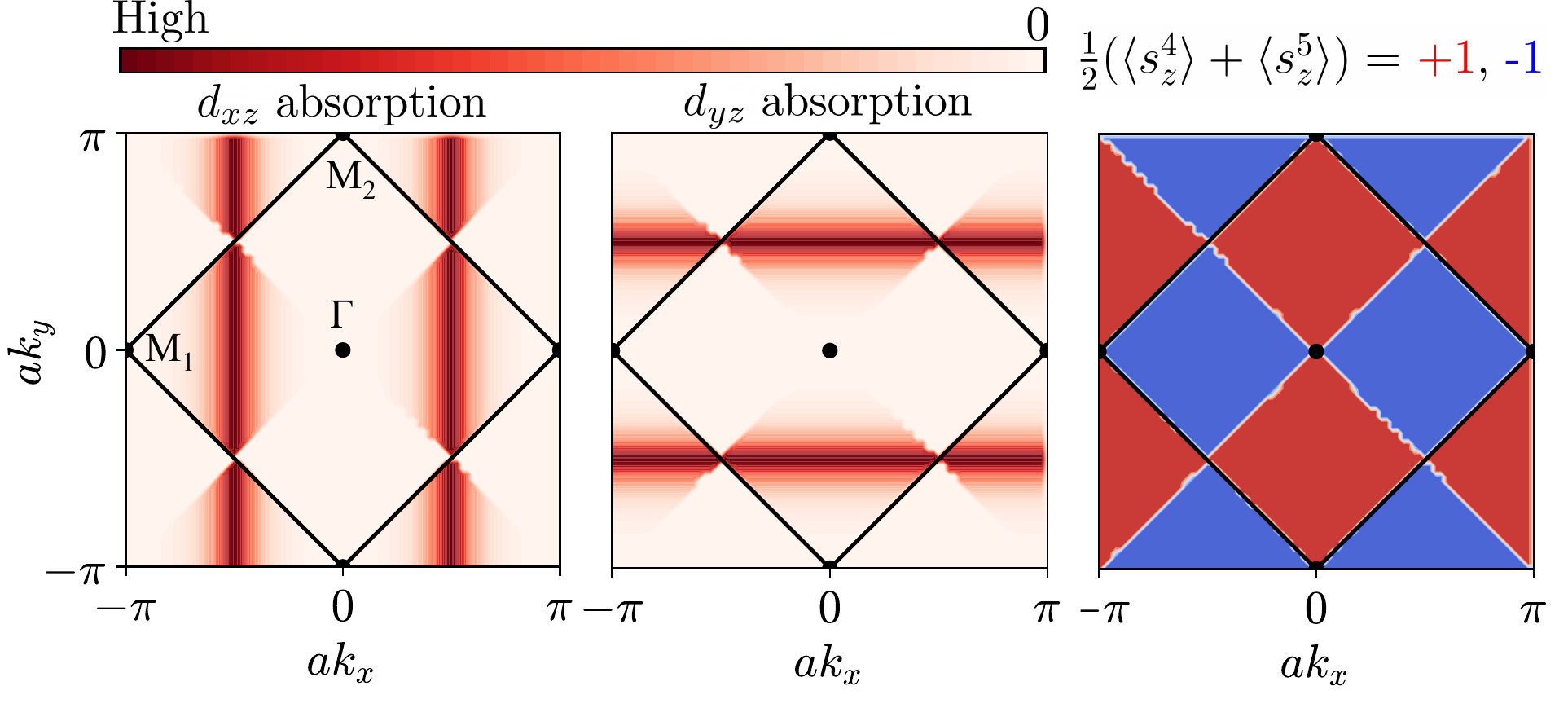}
\caption{Optical matrix elements $|M_{\alpha}^{ji}(\bm{k})|^2$ between bands $i=4, j=5$ in Fig. \ref{fig_fig2}(i). Left (center) plot shows results for light polarized along $\alpha = x$ ($\alpha = y$), which couples to $d_{xz}$ ($d_{yz}$) orbitals. The Brillouin zone and high-symmetry points are highlighted. Right plot shows the momentum-resolved spin polarization of bands 4 and 5 added together. A nonzero value indicates that (i) the bands have the same spin expectation value and hence an electric-dipole optical transition is not forbidden by the spin selection rule and (ii) the spin of the excited electron. The parameters are $V_\pi = 1$, $V_\delta = 0$, $\Delta = 1$ and $m = 1.5$ (in units of $V_\pi$).}
\label{fig_fig3}
\end{figure}

Next, we use these momentum-dependent and spin-selective optical transitions to calculate a clear magneto-optical signature of such \textit{d}-wave ALTM. Specifically, we predict a magnetic linear dichroism in metallic ALTMs. We compute the single-particle absorption spectrum for a given Fermi level $E_F$ and varying photon energy $E_{\hbar\omega}$ as:
\begin{align}\label{eq_absorption}
I_\alpha(E_F,E_{\hbar\omega}) &= \int_{BZ} |M_{\alpha}^{ji}(\bm{k})|^2 \delta(E_j(\bm{k}) - E_i(\bm{k}) - E_{\hbar\omega}) \nonumber \\
& \times f_{FD}(E_i(\bm{k}))(1 - f_{FD}(E_j(\bm{k}))) d\bm{k},
\end{align}
where $E_i(\bm{k})$ and $E_j(\bm{k})$ are the energy of the initial and final states, respectively, and $f_{FD}$ is the Fermi-Dirac distribution function. In absence of a magnetic field, the absorption spectrum of the two light polarizations $\alpha=x,y$ is the same. When a magnetic field \textit{parallel} to the N\'eel vector is applied, a Zeeman shift with the form $\mathcal{H}_Z = \frac{g_s}{2}\mu_B B \sigma_0 \tau_0 s_z$ is introduced in the Hamiltonian (with $g_s$ the spin $g$-factor and $\mu_B$ the Bohr magneton). As seen in Fig. \ref{fig_fig4}(b), this removes the degeneracy between spin up and down bands along the $M_1 - \Gamma$ and $\Gamma - M_2$ paths. %(the $C_4^zT$ symmetry is now broken). 
If the system is metallic, the Fermi surfaces of the two spins will differ implying that the lowest optical transition from each spin-polarized band will not be the same, with spins antiparallel to $B$ being excited at lower energy (see arrows in Fig. \ref{fig_fig4}(b)). As a consequence, the absorption spectra of $x$ and $y$ polarized light will be shifted in energy with respect to one another, thereby producing a magnetic linear dichroism odd in field (i.e. the sign changes upon reversing $B$). We confirm this expectation by numerically computing Eq. \eqref{eq_absorption} for the two polarizations and plotting the results in Fig. \ref{fig_fig4}(a). Importantly, since the spin antiparallel to the magnetic field is the one with the lowest excitation, one can use this measurement to determine the specific orbital-spin coupling: for our model parameters, $x$-polarized light is absorbed at lower energy when $+\vec{B}||\vec{L}$, indicating that $d_{xz}$ ($d_{yz}$) orbitals are coupled to down (up) spins.

\begin{figure}[t!]
    \centering
    \includegraphics[width=0.48\textwidth]{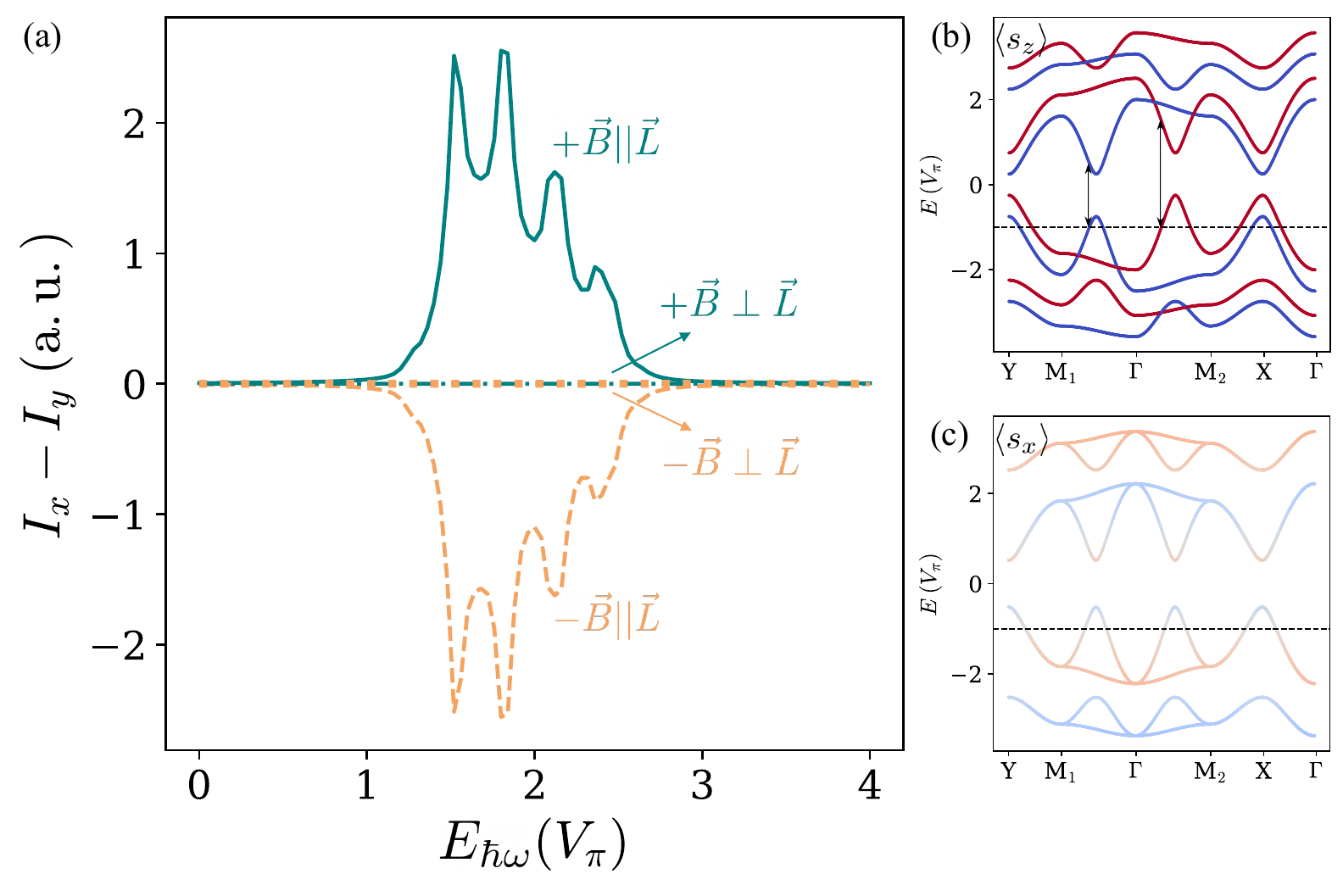}
\caption{(a) Magnetic linear dichroism of a metallic \textit{d}-wave ALTM when a magnetic field $B$ is applied parallel ($\hat{z}$) or antiparallel ($-\hat{z}$) to the N\'eel vector. When the magnetic field is applied perpendicular to it, the dichroism vanishes. (b) Band structure for the case of $+\vec{B}||\vec{L}$ where the color denote the spin polarization along $z$. (c) Same as in (b) but for  $+\vec{B}\perp\vec{L}$ and the color denotes spins along $x$. The Fermi level ($E_F=-1$) is indicated by a dashed line. The parameters are $V_\pi = 1$, $V_\delta = 0.1$, $\Delta = 1$, $m = 1.5$ and $B = 0.25$ (in units of $V_\pi$).}
\label{fig_fig4}
\end{figure}

Another important signature in such magneto-optical measurement appears when the magnetic field is applied \textit{perpendicular} to the N\'eel vector. In this case, the external magnetic field does \textit{not} break the energy degeneracy between the bands in the $M_1 - \Gamma$ and $\Gamma - M_2$ paths (Fig. \ref{fig_fig4}(c)). 
%(The $C_4^zT$ symmetry is still broken, as both pair of bands have the same spin polarization).
As a result, no shift is expected between the different polarizations and thus the magnetic linear dichroism is zero (Fig. \ref{fig_fig4}(a)). Taken together, magneto-optical experiments performed with different magnetic field directions can provide strong evidence of altermagnetism, as well as infer important information such as the nature of the orbital-spin interaction and the N\'eel vector orientation.

\textit{Discussion and outlook --- } In summary, we have inspected the microscopic role of the crystal field in altermagnetism, and through a tight-binding model, we have shown its interplay with the sublattice, orbital and spin degrees of freedom. That allowed us to unveil an orbital-spin locking leading to a clear magnetic linear dichroism signal. Our results broaden the microscopic theoretical understanding of altermagnetism and pave the way for using ALTMs in next-generation opto-spintronic devices.

A notable benefit of our tight-binding approach to predicting optical signatures of altermagnetism is that it moves beyond a symmetry analysis \cite{Kimel2024} by offering microscopic insights into mechanisms behind symmetry-allowed effects, without the need for computationally expensive and material-specific first-principles calculations. We have demonstrated this by predicting a mechanism yielding magnetic linear dichroism in metallic altermagnets (Fig.~\ref{fig_fig4}), which could be observed either by measuring linear dichroism as a function of static magnetic field~\cite{eremenko1987, Higuchi2016}, or by detecting the changes in linear dichroism that are synchronous with an oscillating magnetic field~\cite{sunko2023, donoway2024}. On the other hand, estimating the magnitude of the effect and its photon energy dependence in a specific material is beyond the scope of effective tight-binding models, and requires first-principles calculations, suggesting a powerful synergy between the different theoretical approaches. 
%The empirical comparisons of magnetic linear birefringence with related magneto-optical measurements in transition metal dichalcogenides  \cite{Li2014, MacNeill2015}, suggests that the effect is likely to be observable. 

The demonstrated orbital-spin locking (Fig. \ref{fig_fig3}.) immediately suggests other experiments and offers a perspective on existing ones. For instance, spin-polarized photocurrents  are expected in semiconducting altermagnets: a photoexcitation with linearly polarized light will induce a spin polarization in the conduction band, yielding a spin-polarized current that can be detected through ferromagnetic  electrodes as in nonlocal spin valves \cite{Jedema2001, Vila2020, Luo2017}. Further, the recent experimental results on RuO$_2$~\cite{Weber2024} can be understood within the picture of orbital-spin locking: a lineally polarized light induces an orbital polarization, which in turns induces a non-equilibrium spin polarization,  which is probed with temporal resolution by the magneto-optical Kerr effect \cite{Weber2024, Kato2004, Stamm2017}.

%Finally, some remarks regarding the effects of spin-orbit coupling are in order. It is worth mentioning that the effect of spin-orbit coupling should be small in the optical properties described above. The spin-orbit coupling, at least in the local picture (i.e. $\mathcal{H}_{SOC} \propto \vec{L} \cdot \vec{s}$), should be much weaker than the onsite crystal field energy (Eq. \eqref{eq_CF}) \cite{Occhialini2021}, whose effect is to split the distinct orbitals involved in the angular momentum operator $\vec{L}$ and thus minimizes the effects of spin-orbit. Consequently, the orbital character is expected to still be composed of mainly $d_{xz}$ or $d_{yz}$ orbitals and hence display the same orbital-spin locking properties. Of course, at band crossings with opposite spins, spin-orbit coupling will still mix the bands and be responsible for e.g. anomalous Hall physics \cite{Smejkal2020, Feng2022, Gonzalez2023}, but this should not impact the physics presented here. % but such crossings do not occur along the $\Gamma - M$ path of the bands responsible for the optical properties presented here. 
%\note{Cite several DFT papers saying the orbital character of the bands not only to support our TB but also to say that Lz should be zero or small, so SOC won't do much there}

Our results demonstrate the importance of multi-orbital physics in ALTMs. Therefore, considering the orbital degree of freedom is important not only when altermagnetism arises from correlations \cite{Leeb2024}, but also when the crystal field induces it. Moving forward, it will be exciting to explore whether this magnetic order hosts orbitronics phenomena \cite{Go2021, Jo2024} that could potentially be used to interact with the altermagnetic order parameter. Similarly, the interplay of multi-orbital and crystal field physics with quantum geometric effects recently discovered in ALTMs \cite{Fang2024} could lead to even richer optical properties \cite{Ahn2022}. Finally, we notice that the band structure of ALTMs with magnetic field (Fig. \ref{fig_fig4}(b)) shares similar features with the elusive modified Haldane model \cite{Colomes2018}, suggesting that ALTMs could be used to probe its optical properties \cite{Vila2019}.

\begin{acknowledgments}
M.V is grateful to Omar A. Ashour, Jairo Sinova and Rafael Fernandes for fruitful discussions and critical reading of our manuscript. M.V. and J.E.M. were supported by the Center for Novel Pathways to Quantum Coherence in Materials, an Energy Frontier Research Center funded by the US Department of Energy, Office of Science, Basic Energy Sciences. V.S. received support from the Gordon and Betty Moore Foundation's EPiQS Initiative through Grant GBMF4537. J.E.M. acknowledges a Simons Investigatorship.
\end{acknowledgments}

%********************references************************************************************************
%BibTeX
%\bibliography{bib}


\end{document}